\documentclass[a4paper,11pt]{article}

\usepackage{amsfonts,amssymb}
\usepackage{theorem}

\def\G{\Gamma}

\newcommand{\bea}{\begin{eqnarray}}
\newcommand{\eea}{\end{eqnarray}}

\begin{document}
\

\vskip 2.5 truecm
\Large
\bf
\centerline{Canonical Flow in the Space of Gauge Parameters}
\normalsize \rm

\normalsize
\rm
\vskip 0.9 truecm
\large
\centerline{ 
Andrea Quadri \footnote{E-mail address: 
{\tt andrea.quadri@mi.infn.it}}}

\vskip 0.4 truecm
\normalsize
\centerline{Phys. Dept. University of Milan, 
via Celoria 16, I-20133 Milan, Italy } 
\centerline{and INFN, Sezione di Milano} 

\vskip 0.7  truecm
\normalsize
\bf
\centerline{Abstract}
\rm
\begin{quotation}
\noindent
Gauge dependence of one-particle irreducible (1-PI)
 amplitudes in SU(N) Yang-Mills theory
is shown to be generated by a canonical flow 
with respect to (w.r.t.) the
extended Slavnov-Taylor (ST) identity, induced by the 
transformation of the gauge parameter $\alpha$ 
under the BRST symmetry.
For linear covariant gauges, the analytic expansion  in $\alpha$ 
of 1-PI
amplitudes is given in terms of coefficients
evaluated in the Landau gauge and of derivatives 
w.r.t. $\alpha$ of the generating functional of the flow.
An application to the gauge flow of the 
gluon propagator is considered.
\end{quotation}

\newpage

\section{Introduction}

While physical quantities have to be gauge-invariant,
it sometimes happens that particular gauges are 
computationally more suited than others 
in the study of several properties of gauge theories.

QCD provides a number of examples of this phenomenon.
Just to mention a few,
the computation of the effective action for the Color Glass Condensate
and of the ensuing evolution equations is most easily
carried out in the Light-Cone gauge for the semi-fast 
gluons~\cite{Iancu:2000hn}-\cite{Hatta:2005rn}, so that 
gauge-invariance is not manifest from the beginning.

However it has been
recently proven~\cite{Binosi:2014xua} that  gauge-invariance
of the evolution equations indeed holds  as a consequence
of a suitable Slavnov-Taylor (ST) identity, arising
from the BRST symmetry of QCD
in the presence of the classical fast gluon backgrounds, so that 
any gauge choice for the semi-fast modes can in fact be adopted.

A perhaps more striking example
is the existence of massive solutions of appropriate truncations
 to the QCD Schwinger-Dyson equations~\cite{Aguilar:2008xm,Binosi:2009qm,Fischer:2008uz}, that  has been established 
in the Landau gauge, confirming lattice simulations again carried out in the Landau gauge, both in SU(2)~\cite{Cucchieri:2007md} and in SU(3)~\cite{Bogolubsky:2009dc}.

Moreover, the study of the Kugo-Ojima function is also usually 
formulated in the Landau gauge~\cite{Aguilar:2009pp}.

In the study of the IR properties of QCD it is therefore particularly 
important to try to establish a method, as general as possible, 
in order to ease the comparison of computations carried out 
in different gauges.

On the formal side, it has been known since a long time that 
gauge dependence of amplitudes can be studied algebraically through (generalized) Nielsen identities~\cite{Nielsen:1975fs,Piguet:1984js,Gambino:1999ai}.
Formally these identities can be derived by extending the action
of the BRST differential $s$ 
to the gauge parameters. Their BRST variation is given by classical
anticommuting variables paired into a so-called BRST doublet (i.e.
a pair of variables $u,v$ such that $su = v, sv = 0$).

This technical device bears a close analogy with the algebraic treatment
of gauge theories in the presence of a classical background connection
~\cite{Grassi:1995wr}-\cite{Ferrari:2000yp}, where again the classical background field $\hat A_\mu$
is paired under the BRST differential $s$ with a 
classical anticommuting source $\Omega_\mu$.

We remark that this mathematical structure
indeed arises very naturally if one imposes both the BRST and the antiBRST
symmetry of the underlying gauge theory~\cite{Binosi:2013cea}. 
This requirement  allows to obtain a local antighost equation, valid in any Lorentz-covariant
gauge, by extending the local antighost equation originally derived 
in the Landau gauge~\cite{Grassi:2004yq}.

The resulting extended ST identity turns out to
completely determine the dependence on the background $\hat A_\mu$
of the vertex functional $\G$ through
a canonical transformation w.r.t. the Batalin-Vilkovisky bracket of the theory,
induced by the generating functional $\Psi_\mu \equiv \frac{\delta \G}{\delta \Omega_\mu}$~\cite{Binosi:2011ar,Binosi:2012pd}.

Moreover it has been shown that the solution to the extended ST identity
can be written in terms of a certain Lie series, while naive exponentiation
would fail due to the dependence of $\Psi_\mu$ on the background~\cite{Binosi:2012st}.

In the present paper we will extend these results to the canonical flow
induced by the extended ST identity in the space of gauge parameters.
We will show that such a flow can be derived on the basis of the ST
identity only (so that one can dispose of the equations ensuring
the stability of the gauge-fixing for some particular gauge choices,
like the Nakanishi-Lautrup and the ghost equation~\cite{Piguet:1995er} 
in Lorentz-covariant gauges).

Then we will obtain the explicit form of the Lie series which
gives the expansion of the effective action in powers of the
gauge parameter $\alpha$ (by assuming analyticity in $\alpha$).
The coefficients are given by amplitudes evaluated in the 
theory at $\alpha =0$ (e.g., in the example of Lorentz-covariant gauges,
in terms of Landau gauge amplitudes) plus some contributions
induced by the $\alpha$-dependence of the generating functional of the canonical
flow.

We will then discuss in some detail the gauge flow relating the gluon 
propagator in the Landau and in the Lorentz-covariant gauge. 
The relations derived here are valid in perturbation theory.
It might be tempting to speculate whether they can also be applied
in the non-perturbative regime. This problem
is however well beyond the scope of the present paper, since it involves
the need of a deeper discussion of the analyticity of the gluon propagator
in the gauge parameter around the Landau gauge point at $\alpha=0$, which
might be spoiled beyond perturbation theory.

\medskip
The paper is organized as follows. In Sect.~\ref{sec:cl.act}
we introduce our notations and derive the extended ST identity
for SU(N) Yang-Mills theory. In Sect.~\ref{sec:can.flow}
we construct the canonical flow governing the gauge dependence
of the amplitudes and discuss the role of the gauge dependence
of the generating functional.
In Sect.~\ref{sec:cov} the connection between the Lorentz-covariant
gauges and the Landau gauge is analyzed. Finally in Sect.~\ref{sec:ir}
we discuss an application of the formalism to the gluon propagator.
Conclusions are given in Sect.~\ref{sec:concl}.

\section{Classical Action}~\label{sec:cl.act}

Let us consider pure SU(N) Yang-Mills theory with classical action
\bea
S=-\frac{1}{4g^2} \int d^4x \,  G_{a\mu\nu}^2 \, ,
\label{cl.1}
\eea
with the field strength given by
\bea
G_{a\mu\nu} = \partial_\mu A_{a\nu} - \partial_\nu A_{a\mu} + 
f_{abc} A_{b\mu} A_{c\nu}
\label{cl.2}
\eea
and $f_{abc}$ the SU(N) structure constants.
The inclusion of matter fields does not introduce further complications in the ensuing analysis.

The usual quantization procedure based on the BRST symmetry
requires the introduction in the tree-level vertex functional 
of a gauge-fixing function ${\cal F}_a$ through the
coupling with the Nakanishi-Lautrup multiplier field $b_a$~\cite{NL}:
\bea
S_{g.f.} = - \int d^4x \, b_a {\cal F}_a 
\label{cl.3}
\eea
(the minus sign is inserted for notational convenience).
For the present purposes we do not need to specify the exact form
of the gauge-fixing function ${\cal F}_a$. The only condition
is that it should allow the inversion of the tree-level 2-point functions
in the $A_{a\mu}-b_b$ sector, yielding the 
tree-level propagators for the gauge and Nakanishi-Lautrup multiplier fields.
${\cal F}_a$ might also depend on some parameters $\lambda_i$.
For instance, one might choose
\bea
{\cal F}_a = (1 - \lambda) \partial^\mu A_{a\mu} + \lambda~ \partial^i A_i
\label{cl.4}
\eea
interpolating between the Lorentz-covariant gauge ($\lambda =0$) and 
the Coulomb gauge ($\lambda=1$).
Another example is the Slavnov-Frolov regularization of the Light Cone gauge~\cite{Slavnov:1987yh}
\bea
{\cal F}_a = A_- + \lambda \partial_- A_-
\eea
where $A_- = A_0 - A_3$ and $\partial_- = \partial_0 - \partial_3$.
Green functions are evaluated at $\lambda \neq 0$ and then one takes
the limit $\lambda \rightarrow 0$.

Gauge invariance lost after the gauge-fixing procedure
is promoted to full BRST symmetry by adding 
to the classical action
both the gauge-fixing and the ghost-dependent terms
\bea
S_{g.f.+ gh} & = & s \int d^4x \, \bar c_a \left ( \frac{\alpha}{2} b_a - {\cal F}_a \right ) \nonumber \\
 & = & \int d^4x \, \left ( \frac{\alpha}{2} b_a^2 - b_a {\cal F}_a 
+ \bar c_a s {\cal F}_a \right ) \, ,
\label{cl.gfANDgh}
\eea
where the nilpotent BRST differential $s$ acts as follows. 
On the gauge field it equals the gauge transformation, upon replacement of the gauge parameters with the ghost fields $c_a$
\bea
s A_{a\mu} \equiv D_\mu c_a = \partial_\mu c_a + f_{abc} A_{b\mu} c_c \, ,
\label{brst.a}
\eea
where $D_\mu c_a$ is the covariant derivative of the ghost field.
The transformation of the ghost is in turn dictated by the nilpotency of $s$, i.e.
\bea
s c_a = -\frac{1}{2} f_{abc} c_b c_c \, .
\label{brst.c}
\eea
The antighost $\bar c_a$ and the Nakanishi-Lautrup multiplier field $b_a$ form a BRST doublet~\cite{Barnich:2000zw,Quadri:2002nh}, i.e.
\bea
s \bar c_a = b_a \, , ~~~~~ s b_a = 0 \, .
\eea
The parameter $\alpha$ reduces to the usual gauge parameter for Lorentz-covariant gauges when ${\cal F}_a = \partial A_a$.

\subsection{Slavnov-Taylor Identity}

Since the BRST variations of the gauge and ghost fields
in Eqs.(\ref{brst.a}) and (\ref{brst.c}) are non-linear in the quantum fields, their renormalization
requires the introduction of external sources known as antifields~\cite{Gomis:1994he}.
They are coupled to the BRST variation of the corresponding fields as follows
\bea
S_{a.f.} = \int d^4x \, \left ( A^*_{a\mu} s A_{a\mu} - c^*_a s c_a \right ) \,.
\label{cl.af}
\eea
The minus sign in front of $c^*_a s c_a$ is introduced for consistency
with the Batalin-Vilkovisky (BV) bracket conventions of~\cite{Gomis:1994he}.

Then the tree-level vertex functional
\bea
\G^{(0)} = S + S_{g.f.+gh} + S_{a.f.}
\eea
obeys the ST identity~\cite{Slavnov:1972fg,Taylor:1971ff}
\bea
{\cal S}(\G^{(0)}) \equiv \int d^4x \, \Big (
\frac{\delta \G^{(0)}}{\delta A^*_{a\mu}} 
\frac{\delta \G^{(0)}}{\delta A_{a\mu}} 
-
\frac{\delta \G^{(0)}}{\delta c^*_{a}} 
\frac{\delta \G^{(0)}}{\delta c_{a}}
+
b_a \frac{\delta \G^{(0)}}{\delta \bar c_a} \Big ) = 0 \, .
\label{st.cl}
\eea
Notice that the linearity of the BRST transformation of the antighost
$\bar c_a$ does not strictly require the introduciton of an antifield
for $\bar c_a$.

The  ST identity in Eq.(\ref{st.cl}) holds irrespectively of the particular form of the
gauge-fixing function ${\cal F}_a$ chosen. 
For some specific choices of the latter (e.g.  linear covariant gauges or the Landau gauge) further identities arise, like the equation for the 
$b$-field and the ghost equation~\cite{Piguet:1995er}. However, 
%for the sake of generality, 
we will not rely on these identities in the following discussion.

\subsection{BRST Variation of the Gauge Parameters}

It has been known since a long time~\cite{Piguet:1984js} that 
one can extend the BRST symmetry to act on the gauge parameters 
in such a way to derive an extended ST identity, leading to the
so-called Nielsen identities~\cite{Nielsen:1975fs,Piguet:1984js}.
I.e. one defines 
\bea
s \lambda_i = \theta_i \, , \quad s \theta_i = 0 \, , 
\quad s \alpha = \theta \, \quad s \theta = 0 \, .
\label{brst.par}
\eea
Under the extended BRST symmetry, $S_{g.f. + gh}$ in Eq.(\ref{cl.gfANDgh}) 
receives an additional contribution
\bea
S_{g.f.+ gh} & = & s \int d^4x \, \bar c_a \left ( \frac{\alpha}{2} b_a - {\cal F}_a \right ) \nonumber \\
 & = & \int d^4x \, \left ( \frac{\alpha}{2} b_a^2 - b_a {\cal F}_a 
+ \bar c_a s {\cal F}_a + \frac{\theta}{2} \bar c_a b_a \right ) \nonumber \\
 &   & + \int d^4x \,  \bar c_a
\left ( \frac{\partial {\cal F}_a}{\partial \lambda_i} \theta_i + 
 \frac{\partial {\cal F}_a}{\partial \alpha} \theta 
\right )
\label{cl.gfANDgh.ext}
\eea
and the tree-level classical action fulfills the extended ST identity
\bea
{\tilde{\cal S}}(\G^{(0)}) = 
\sum_i \theta_i \frac{\partial \G^{(0)}}{\partial \lambda_i} 
+ \theta \frac{\partial \G^{(0)}}{\partial \alpha} +
{\cal S}(\G^{(0)}) = 0 \, .
\label{ext.sti}
\eea
For non-anomalous theories this equation holds for the full vertex functional~$\G$:
\bea
{\tilde{\cal S}}(\G) = 
\sum_i \theta_i \frac{\partial \G}{\partial \lambda_i} 
+ \theta \frac{\partial \G}{\partial \alpha} +
{\cal S}(\G) = 0 \, .
\label{ext.sti.full}
\eea
By taking a derivative w.r.t. $\theta$ and then setting $\theta, \theta_i$ equal to zero one obtains the following Nielsen identity
\bea
\left . 
 \frac{\partial \G}{\partial \alpha}
\right |_{\theta=\theta_i=0}
 & = & -
\int d^4x \, \Big (
\frac{\delta^2 \G}{\partial \theta \delta A^*_{a\mu}} 
\frac{\delta \G}{\delta A_{a\mu}} 
-
\frac{\delta \G}{\delta A^*_{a\mu}} 
\frac{\delta^2 \G}{\partial \theta \delta A_{a\mu}} 
\nonumber \\
& & \left . -
\frac{\delta^2 \G}{\partial \theta \delta c^*_{a}} 
\frac{\delta \G}{\delta c_{a}}
-
\frac{\delta \G}{ \delta c^*_{a}} 
\frac{\delta^2 \G}{\partial \theta \delta c_{a}}
+
b_a \frac{\delta^2 \G}{\partial \theta \delta \bar c_a} \Big ) 
\right |_{\theta=\theta_i=0}
\, .
\label{nielsen.id}
\eea
A similar equation holds for the derivative of $\G$ w.r.t.
$\lambda_i$, once one takes a derivative of the extended ST
identity in eq.(\ref{ext.sti.full}) w.r.t. $\theta_i$.

\section{Canonical Flow for Gauge Parameters}\label{sec:can.flow}

There is a close formal analogy between Eq.(\ref{ext.sti.full}) and
the extended ST identity controlling the dependence
on a background field configuration $\widehat A_{\mu}$~\cite{Grassi:1995wr}-\cite{Ferrari:2000yp}.
This analogy relies on the fact that both the gauge parameters
and the background configurations form  BRST doublets
(i.e. a couple of variables $u,v$, such that $su = v \, , s v =0$)
together with their BRST partners.

Assuming analyticity in the background field configuration,
the solution to the extended ST identity can be obtained
by a suitable Lie series~\cite{Binosi:2012st}
that allows to express
all the coefficients in the expansion in powers of $\widehat A_\mu$
in terms of Green functions evaluated at zero background.
The failure of naive exponentiation and the need to use a
Lie series arises from the dependence of the generating
functional, controlling the background dependence, on the 
background itself.

The same technique can be used to obtain a Lie series for the
expansion of the vertex functional in powers of $\alpha$
(or $\lambda_i$) in terms of amplitudes evaluated at $\alpha = 0$
(or $\lambda_i=0)$. Again one assumes analyticity in the parameter
$\alpha$ (or $\lambda_i$) one is considering.

For that purpose it is convenient to rewrite the extended
ST identity within the BV formalism~\cite{Gomis:1994he}.
Hence one introduces an antifield $\bar c^*_a$ for the antighost
$\bar c_a$ as well as the antifield $b^*_a$ 
for the Nakanishi-Lautrup field $b_a$.
$\bar c^*_a$ is coupled to $b_a$ in the classical action, while
$b^*_a$ does not enter into $\G^{(0)}$ (since $s b_a = 0$).

Then one defines the BV bracket as follows (we use only left derivatives)
\bea
&& 
\!\!\!\!\!\!\!\!\!\!\!\!\!\!\!\!\!\!\!\!\!\!\!
\{ X, Y \}  =  \int d^4x \, \sum_\phi \Bigg [
(-1)^{\epsilon_\phi (\epsilon_X+1)} \frac{\delta X}{\delta \phi} 
\frac{\delta Y}{\delta \phi^*} - (-1)^{\epsilon_{\phi^*} (\epsilon_X +1 )}
\frac{\delta X}{\delta \phi^*}\frac{\delta Y}{\delta \phi} \Bigg ] \, .
\label{bv.bracket}
\eea
The sum runs over the fields $\phi = (A_{a\mu}, c_a, \bar c_a, b_a)$
and the corresponding antifields $\phi^* = (A^*_{a\mu}, c^*_a, \bar c_a^*, b_a^*)$. $\epsilon_\phi, \epsilon_{\phi^*}$ are the statistics of the field $\phi$ and the antifield $\phi^*$. $\epsilon_X$ is the statistics of the functional $X$.

Then the extended ST identity (\ref{ext.sti.full}) can be written as
\bea
\tilde{\cal S}(\G) = 
\sum_i \theta_i \frac{\partial \G}{\partial \lambda_i} 
+ \theta \frac{\partial \G}{\partial \alpha} + \frac{1}{2} \{ \G, \G \} = 0
\, .
\label{st.bv}
\eea
By taking a derivative w.r.t. $\theta$ (but the argument goes
in the same way if one takes a derivative w.r.t. $\theta_i$) one finds
\bea
\left . \frac{\partial \G}{\partial \alpha} \right |_{\theta = \theta_i =0} 
= - \left . \{ \frac{\partial \G}{\partial \theta}, \G \} 
\right |_{\theta = \theta_i =0} \, .
\label{eq.alpha}
\eea
This equation shows that the derivative of the vertex functional
w.r.t. $\alpha$ is obtained by a canonical transformtion 
(w.r.t. the BV bracket) induced by the 
generating functional $\Psi \equiv \frac{\partial \G}{\partial \theta}$.
Since the latter in general depends on $\alpha$, one cannot
solve Eq.(\ref{eq.alpha}) by simple exponentiation and one needs
to make recourse to a Lie series.

For that purpose, one introduces the operator
\bea
\Delta_\Psi = \{ \cdot, \Psi \} + \frac{\partial}{\partial \alpha} \, .
\label{Delta}
\eea
Then the vertex functional $\G$ is given by the following Lie
series~\cite{Binosi:2012st}
\bea
\G = \sum_{n \geq 0} \frac{1}{n!} \alpha^n [\Delta_\Psi^n \G_0]_{\alpha = 0}
\label{Lie.series}
\eea
where $\G_0$ is the vertex functional at $\alpha=0$.
Notice that one must afterwards take the limit $\alpha \rightarrow 0$
(although the operator $\Delta_\Psi$ is applied on the functional
$\G_0$, which is $\alpha$-independent) since a residual $\alpha$-dependence
may arise (and in general indeed arises) from the differentiation w.r.t. $\alpha$ 
of the generating functional $\Psi$.

We also remark  that Eq.(\ref{Lie.series}) holds irrespectively of the
form of the gauge-fixing function ${\cal F}_a$ (and in particular 
is independent of the existence of a $b$-equation
and of a ghost equation, guaranteeing the stability of the gauge-fixing
in certain classes of gauge~\cite{Piguet:1995er}).

\section{Lorentz-covariant Gauges}\label{sec:cov}

Let us illustrate the above formalism in the simple example 
of the Lorentz-covariant gauge, i.e. let us choose
\bea
{\cal F}_a = \partial A_a \, .
\label{Lorentz.cov}
\eea
Then one gets
\bea
S_{g.f.+gh} & = & \int d^4 x \, \Big ( \frac{\alpha}{2} b_a^2 - b_a \partial A_a + \bar c_a \partial^\mu (D_\mu c)_a + \frac{\theta}{2} \bar c_a b_a \Big ) \, .
\label{gf.Lor.cov}
\eea
The propagators are 
\bea
&& \Delta_{A^a_\mu A^b_\nu} = - i \delta^{ab} 
\Big ( \frac{1}{p^2} T^{\mu\nu} + \frac{\alpha}{p^2} L^{\mu\nu} \Big ) \, , 
\qquad \Delta_{b_a A_{b\mu}} = - \delta^{ab} \frac{p^\mu}{p^2} \, , \nonumber \\
&& 
\Delta_{b_a b_b} = 0 \, , 
\qquad \Delta_{c_a \bar c_b} = \delta_{ab} \frac{i}{p^2} \, .
\label{prop}
\eea
$T^{\mu\nu} = g^{\mu\nu} - \frac{p^\mu p^\nu}{p^2}$ is the transverse
projector, $L^{\mu\nu} = \frac{p^\mu p^\nu}{p^2}$ is the longitudinal one.

For Lorentz covariant gauges the $b$-equation and the ghost equation
hold:
\bea
&& \frac{\delta \G}{\delta b_a} = \alpha b_a - \partial A_a \, ,
\nonumber \\
&&  \frac{\delta \G}{\delta \bar c_a} = 
\partial^\mu \frac{\delta \G}{\delta A^*_{a\mu}} - \frac{\theta}{2} b_a \, .
\label{stab.eqs}
\eea
The first of the above equations implies that the $b$-dependence is confined at tree level. The second equation in turn implies that at higher orders $n \geq 1$
$\G$ can depend on $\bar c_a$ only through the 
combination
\bea
{\tilde A}^*_{a\mu} = A^*_{a\mu} - \partial_\mu \bar c_a \, .
\label{new.af}
\eea
By redefining the antifield $A^*_{a\mu}$ according to the above equation
and by introducing the reduced functional 
$\tilde \G = \G - \int d^4x \, \frac{\alpha}{2} b_a^2 +
\int d^4x \,   b_a \partial A_a$, the BV bracket can be restricted
to the variables $(A_{a\mu},{\tilde A}^*_{a\mu})$ and 
$(c_a, c^*_a)$ and the flow equation reads
\bea
\left . \frac{\partial \tilde \G}{\partial \alpha} 
\right |_{\theta=\theta_i=0}  = 
- \int d^4x \Bigg [ 
\frac{\delta \Psi}{\delta A_{a\mu}} 
\frac{\delta \tilde \G}{\delta {\tilde A}^*_{a\mu}} +
\frac{\delta \Psi}{\delta {\tilde A}^*_{a\mu}}
\frac{\delta \tilde \G}{\delta A_{a\mu}} 
- \frac{\delta \Psi}{\delta c_a} 
\frac{\delta \tilde \G}{\delta c^*_a} -
\frac{\delta \Psi}{\delta c^*_a}
\frac{\delta \tilde \G}{\delta c_a}  \Bigg ] \, .
\label{flow.diff}
\eea 
Since we will only use $\tilde \G$  in what follows, we will simply write $\G$ for $\tilde \G$.
The Lie operator $\Delta_\Psi$  is
\bea
\Delta_\Psi(X) &  =  & \{ X, \Psi \} + \frac{\partial X}{\partial \alpha} 
\nonumber \\
& = &
\int d^4x \Bigg [ 
\frac{\delta X}{\delta A_{a\mu}} 
\frac{\delta \Psi}{\delta {\tilde A}^*_{a\mu}} +
\frac{\delta X}{\delta {\tilde A}^*_{a\mu}}
\frac{\delta \Psi}{\delta A_{a\mu}} 
- \frac{\delta X}{\delta c_a} 
\frac{\delta \Psi}{\delta c^*_a} -
\frac{\delta X}{\delta c^*_a}
\frac{\delta \Psi}{\delta c_a} \Bigg ] \nonumber \\
& &  + 
\frac{\partial X}{\partial \alpha} \, .
\eea

By Eq.(\ref{gf.Lor.cov}) we see that 
at tree-level $\Psi$ reduces to
\bea
\Psi = \int d^4x \, \frac{1}{2} \bar c_a b_a + O(\hbar) \, .
\label{psi}
\eea
In the present case $\G_0$ in Eq.(\ref{Lie.series}) is the
vertex functional of Yang-Mills theory in the Landau gauge.
Eq.(\ref{Lie.series}) then allows one to express
the coefficients of the $\alpha$-expansion of 1-PI amplitudes
in the Lorentz-covariant gauge in terms of 1-PI Landau gauge
amplitudes plus an $\alpha$-dependent contribution, arising from the generating
functional $\Psi$.

The coefficient $\G_1$ is obtained according to Eq.(\ref{Lie.series})
by applying $\Delta_\Psi$ once on $\G_0$. Since $\G_0$ does not
depend on $\alpha$, one obtains
\bea
&& \!\!\!\!\!\!\!\!\!\!\!\!\!\!\!\!
\G_1  =  \int d^4x \, \left . \Big ( 
\frac{\delta \G_0}{\delta A_{a\mu}} \frac{\delta \Psi}{\delta
{\tilde A}^*_{a\mu}}
+ 
 \frac{\delta \G_0}{\delta
{\tilde A}^*_{a\mu}}
\frac{\delta \Psi}{\delta A_{a\mu}}
 %\right . \nonumber \\
%& & \left . 
- \frac{\delta \G_0}{\delta c_a} \frac{\delta \Psi}{\delta c^*_a}
- \frac{\delta \G_0}{\delta c^*_a} \frac{\delta \Psi}{\delta c_a} \Big )
\right |_{\alpha=0} \, .
\label{Gamma.1}
\eea
This equation expresses the linear approximation to $\G$ in powers
of the gauge parameter $\alpha$, in terms of amplitudes evaluated in the Landau gauge.

\medskip
Let us now go on by computing $\G_2$, defined by 
\bea
\G_2 = \left . \frac{\partial^2 \G}{\partial \alpha^2} \right |_{\alpha=0} \, .
\label{Gamma.2}
\eea
According to Eq.(\ref{Lie.series}), this is obtained by applying
$\Delta_\Psi$ twice on $\G_0$ and then setting $\alpha=0$.
Now we get two pieces: the first one again only contains
amplitudes in the Landau gauge and can be written concisely
as $\left . \{ \Psi, \{ \Psi, \G_0 \} \} \right |_{\alpha=0}$.
This is the term associated with naive exponentiation.
However, there is also a contribution arising from the derivative of $\Psi$
w.r.t. $\alpha$, so that the full $\G_2$ reads
\bea
\G_2 & = & \left . \{ \Psi, \{ \Psi, \G_0 \} \}  \right |_{\alpha=0} + \nonumber \\
     &   & + \int d^4x \, \Big [ \frac{\delta \G_0}{\delta A_{a\mu}}
\frac{\delta^2 \Psi}{\partial \alpha \delta {\tilde A}^*_{a\mu}}
+ \frac{\delta \G_0}{\delta  {\tilde A}^*_{a\mu}} 
\frac{\delta^2 \Psi}{\partial \alpha \delta A_{a\mu}} \nonumber \\
& & \qquad \quad \quad - \frac{\delta \G_0}{\delta c_a} \frac{\delta^2 \Psi}{\partial \alpha \delta c^*_a} - \frac{\delta \G_0}{\delta c_a^*} \frac{\delta^2 \Psi}{\partial \alpha \delta c_a} \Big ]_{\alpha=0} \, .
\eea

\section{Gauge Dependence of the Gluon Propagator}\label{sec:ir}

As an example, let us consider in the perturbative regime 
how one can derive the solution to the gauge evolution equation
for the transverse part of the 
gluon propagator. For that purpose 
we introduce the transverse and longitudinal form factors according to
\bea
\Delta_{A^a_\mu A^b_\nu} = -i \delta^{ab} \Big ( 
\Delta_T(p^2) T^{\mu\nu} + \Delta_L(p^2) L^{\mu\nu} \Big )  \, .
\label{gluon.1}
\eea
The relevant quantity is $\Delta_T(p^2)$. 
By taking two derivatives of Eq.(\ref{flow.diff}) w.r.t. $A_{b_1\nu_1}, A_{b_2 \nu_2}$ and 
then setting all fields and external sources to zero we obtain
\bea
\frac{\partial \G_{A_{b_1 \nu_1} A_{b_2 \nu_2}}}{\partial \alpha} 
= - \int d^4x \,\Big [  \G_{\theta {\tilde A}^*_{a\mu} A_{b_1 \nu_1}} \G_{A_{b_2 \nu_2} A_{a\mu}}
+ \G_{\theta {\tilde A}^*_{a\mu} A_{b_2 \nu_2}} \G_{A_{b_1 \nu_1} A_{a\mu}} \Big ] \, .
\label{2pt}
\eea
In the above equation we have used the short-hand notation where lowstair letters denote functional differentiation w.r.t. that argument and
it is understood that in the end one sets all fields $\Phi$ 
and external sources $\Phi^*,\theta,\theta_i$ equal to zero. For instance
\bea
\G_{A_{b_1 \nu_1} A_{b_2 \nu_2}} \equiv \left. \frac{\delta^2 \G}{\delta A_{b_1 \nu_1} 
\delta A_{b_2 \nu_2}} \right |_{\Phi=\Phi^*=\theta=\theta_i=0} \, .
\label{1pi.fnct}
\eea
Let us introduce transverse and longitudinal form factors for 
the 1-PI functions involved, namely (in the Fourier space)
\bea
&& \G_{A_{b_1 \nu_1} A_{b_2 \nu_2}} = \delta_{b_1 b_2} \Big ( G^T T_{\mu\nu} + G^L L_{\mu\nu} \Big ) \, , \nonumber \\
&& \G_{\theta {\tilde A}^*_{a\mu} A_{b \nu}} = \delta_{a b} \Big ( R^T T_{\mu \nu}
+ R^L L_{\mu\nu} \Big ) \, .
\label{form.fact}
\eea
Then by applying the transverse projector to Eq.(\ref{2pt}) one gets
\bea
\frac{\partial G^T}{\partial \alpha} = - 2 R^T G^T \, .
\label{diff.eq}
\eea
Let us denote by $G^T_0$ the form factor in the Landau gauge.
Then by integrating Eq.(\ref{diff.eq}) one gets
\bea
G^T = \exp \Big ( - \int_0^\alpha 2 R^T ~ d\alpha' \Big ) G^T_0
\eea
and therefore for the transverse part of the gluon propagator
\bea
\Delta^T =\exp \Big ( \int_0^\alpha ~ 2 R^T ~ d\alpha' \Big ) \Delta ^T_0 \, .
\label{rel.prop}
\eea

On the other hand, by Eq.(\ref{rel.prop}) the following ratio 
\bea
r =  \exp \Big ( - \int_0^\alpha 2 R^T ~ d\alpha' \Big ) \frac{\Delta^T}{\Delta^T_0}
\label{ratio}
\eea
must be equal to one (and therefore gauge-independent).

\medskip
While these results are valid in the perturbative
expansion, their extension beyond perturbation theory 
 is a
subtle issue whose study is well beyond the scope of this work.

Several computations in the Landau gauge based on 
Schwinger-Dyson equations have indeed identified
a scaling solution with $\Delta_T(0)=0$~\cite{Alkofer:2000wg} and a
 decoupling one, with $\Delta_T(0) > 0$~(see Refs.~\cite{Aguilar:2008xm,Binosi:2009qm} and references therein). 

It is therefore important to study what happens to
these classes of solutions under a gauge variation, e.g.
in order to compare the evolution with existing lattice results
at $\alpha \neq 0$~\cite{Cucchieri:2011pp}.

If one were allowed to 
take the IR limit in both sides of the Eq.(\ref{rel.prop}),
$\Delta^T_0(0)=0$ would imply that
$\Delta^T(0)$ is also equal to zero. 
I.e. if a solution to the
QCD Schwinger-Dyson equations is of the scaling type in the Landau gauge,
it would  also be scaling in a Lorentz-covariant gauge.
Moreover, for massive solutions 
the sign of $\Delta^T(0)$ would be gauge-independent, as a consequence
of Eq.(\ref{rel.prop}).

However the validity of Eq.(\ref{rel.prop}) beyond
perturbation theory is questionable.
In particular, the presence of IR divergences in
the explicit non-perturbative evaluation of the form factor $R^T$ 
might destroy the validity of the assumption that 
the amplitudes are analytic around the Landau gauge point $\alpha=0$.
In this case the Lie series solution in Eq.(\ref{Lie.series}) 
cannot be any more used to reconstruct the vertex functional
in a gauge $\alpha \neq 0$.

\section{Conclusions}\label{sec:concl}

The existence of a canonical flow in the space of gauge parameters
and the related solution in terms of a Lie series provide a way
to compare results in different gauges within an algebraic framework
that is bound to hold even beyond perturbation theory (as far as 
the ST identity is valid).

The dependence of the generating functional of the canonical flow
on the gauge parameter prevents to get the full solution
by a naive exponentiation. Such a solution can be expressed
through an appropriate Lie series, in close analogy to the
solution of the extended ST identity in the presence of
a background gauge connection.

Knowing such a Lie series eases the comparison
between computations carried out in different gauges. In the
simplest example of the 2-point gluon function, a closed formula
interpolating between the Landau and the Lorentz-covariant gauge
can be obtained, under the assumption that analyticity in the 
gauge parameter around $\alpha=0$ holds.

\section*{Acknowledgments}

A critical reading of the manuscript by D.\ Binosi is gratefully acknowledged.

\end{document}